\documentclass[prb,twocolumn,showpacs,floatfix,amsmath,amssymb,aps]{revtex4-1}
\usepackage{dcolumn}
\usepackage{xcolor}
\usepackage[utf8]{inputenc}
\usepackage{amsmath}
\usepackage{amssymb}
\usepackage{dsfont}
\usepackage{physics}
\usepackage{color}
\usepackage{graphicx}
\usepackage{mathtools}
\usepackage{overpic}
\usepackage[toc,page]{appendix}
\usepackage{chemfig}
\usepackage{ccycle}
\usepackage{comment}
\usepackage{textcomp}
\usepackage{gensymb}
\usepackage{graphics}
\usepackage{epsfig}
\usepackage{subfigure}
\usepackage{siunitx}
\usepackage[T1]{fontenc}
\usepackage{soul}
\usepackage{mathrsfs}

\usepackage{hyperref}
\hypersetup{
    colorlinks=true,
    linkcolor=cyan,
    urlcolor=cyan,
    citecolor=cyan,
    }

\newcommand{\shortOper}[1]{\!#1\!}

\begin{document}
\preprint{APS}

\title{Wannier interpolation of reciprocal-space periodic and non-periodic matrix elements in the optimally smooth subspace}

\author{Giulio Volpato}
\affiliation{Department of Physics, University of Trento, Via Sommarive 14, 38123 Povo, Italy}

\author{Stefano Mocatti}
\affiliation{Department of Physics, University of Trento, Via Sommarive 14, 38123 Povo, Italy}

\author{Giovanni Marini}
\affiliation{Department of Physics, University of Trento, Via Sommarive 14, 38123 Povo, Italy}

\author{Matteo Calandra} 
\affiliation{Department of Physics, University of Trento, Via Sommarive 14, 38123 Povo, Italy}

\begin{abstract}

Maximally localized Wannier functions use the gauge freedom of Bloch wavefunctions to define the optimally smooth subspace with matrix elements that depend smoothly on crystal momentum. The associated Wannier functions are real-space localized, a feature often used to Fourier interpolate periodic observables in reciprocal space on ultradense momentum grids. However, Fourier interpolation cannot handle non-periodic quantities in reciprocal space, such as the oscillator strength matrix elements, which are crucial for the evaluation of optical properties. We show that a direct multidimensional interpolation in the optimally smooth subspace yields comparable accuracy with respect to Fourier interpolation at a similar or lower computational cost. This approach can also interpolate and extrapolate non-periodic observables, enabling the calculation of optical properties on ultradense momentum grids. Finally, we underline that direct interpolation in the optimally smooth subspace can be employed for periodic and non-periodic tensors of any order without any information on the position of the Wannier centers in real space.

\end{abstract}

\maketitle
\section{Introduction}\label{sec:Intro}

Maximally Localized Wannier functions (MLWF) are an efficient and widespread technique to obtain a tight binding
representation of the Hamiltonian on a basis that is generally localized in real space both for metals and insulators \cite{PhysRevB.56.12847,PhysRevB.65.035109,RevModPhys.84.1419}.
A notable consequence of this approach is that the interpolation of electronic structure, Berry phase, Hall conductivity \cite{PhysRevB.75.195121} and electron-phonon interaction \cite{PhysRevB.76.165108,PhysRevB.82.165111} on ultradense crystal-momentum grids can be carried out at a marginal cost compared with a first principles calculation.

Interpolation is, however, possible only for observables with the periodicity of the Brillouin zone (BZ) in reciprocal space; there are currently no techniques to Wannier interpolate or extrapolate non-periodic observables. A prominent example is the oscillator strength that rules properties such as optical absorption and photoluminescence. The convergence of optical spectra with Brillouin zone sampling is extremely slow. Moreover, a gauge invariant spectrum requires oscillator strengths calculated with the same electronic wavefunctions used for the electron-phonon matrix elements \cite{doi:10.1021/acs.nanolett.4c00669, PhysRevLett.125.107401, Chan2023}. To exploit the full power of MLWF interpolation it is desirable to develop a strategy to interpolate or extrapolate non-periodic observables in reciprocal space.

To understand the reason for the outlined limitation, it is worth recalling some properties of MLWF. The key idea of MLWF is to exploit the invariance of the observables under rotation of the Bloch functions in a given submanifold of electronic states (either a composite set of bands \cite{PhysRevB.56.12847} or the set of states resulting from a disentanglement procedure \cite{PhysRevB.65.035109}). As a Bloch functions rotation leads to a new set of Wannier functions with different localization properties, it is possible to select accurately the rotational matrix to obtain Wannier functions with the smallest spread in real space. It is known that the dependence on crystal momentum of Bloch functions yielding localized Wannier functions displays a certain degree of smoothness and the smoothness increases with the localization of the Wannier functions in real space \cite{PhysRev.135.A685,PhysRev.115.809, PhysRevLett.98.046402}. On the contrary, Bloch functions resulting from a first principles calculation have a non-continuous, highly non-analytical dependence on crystal momentum. For this reason, determining localized Wannier functions is equivalent to find a set of Bloch functions forming the so-called optimally smooth subspace. In this subspace, any interpolation with respect to the crystal
momentum is possible as its dependence is smooth. On the contrary, the interpolation of any observable is unfeasible in the Bloch function basis resulting directly from a first principles calculation.

Currently, Wannier interpolation is performed through the following steps: (i) the observable is calculated from first principles on a coarse crystal momentum grid; (ii) the transformation to the optimally smooth subspace is performed, yielding an observable written in the basis of Bloch functions that are smooth with respect to the crystal momentum; (iii) the observable is transformed to real space in the basis of the MLWF and, finally, (iv) Fourier interpolation on ultradense crystal momentum grids is carried out.
Steps (iii) and (iv) require that the observable is periodic in reciprocal space, thus hindering the interpolation of non-periodic observables.

In this work, we demonstrate that a different strategy can be considered to interpolate periodic and non-periodic observables in reciprocal space. The idea is to perform a direct multidimensional interpolation in the optimally smooth subspace (i.e., replacing steps (iii) and (iv)) without performing any transformation to real space in the basis of MLWF. As reciprocal space periodicity is not required, non-periodic quantities can be interpolated on ultradense crystal momentum grids provided that their values are known on coarse grids in the interpolation regions (i.e., the second, third, fourth shells around the first Brillouin zone).
Observables periodic in reciprocal space are also interpolated via this technique at a cost comparable to or lower than the standard approach. Finally, we underline that the multidimensional interpolation in the optimally smooth space preserves the smoothness of the phase of the electronic wavefunctions and can thus be used to calculate geometrical objects such as the Berry phase.

We apply this method to the calculation of electron-phonon coupling in MgB$_2$ and single-layer electron-doped MoS$_2$ and of oscillator strengths on the latter compound. In all considered cases, we find that for periodic quantities in reciprocal space, multidimensional direct interpolation is slightly more accurate than standard interpolation while maintaining comparable or lower computational overhead. For non-periodic matrix elements, it leads to accurate interpolated quantities that are unattainable at a reasonable cost using Fourier interpolation.

The paper is structured as follows. In Sec. II we recall all the fundamental concepts related to the interpolation in the smooth subspace, in Sec. III we introduce the new multidimensional interpolation scheme in the optimally smooth subspace, in Sec. IV we demonstrate some exemplar applications, in Sec. V we discuss the new method in light of the results. Finally, in Sec.VI we draw our conclusions.

\section{Wannier functions and the optimally smooth subspace}\label{sec:WannFunc}

For the sake of simplicity, we consider a composite set of bands, i.e., a set of $N_{w}$ bands isolated from all the others. In an insulator, it is always possible to identify such a set of bands. 
In metals, a disentanglement procedure needs to be performed  \cite{PhysRevB.65.035109} to isolate an appropriate subset of target bands behaving similarly to a composite set. 

The single-particle Kohn-Sham Bloch functions  $|\psi_{{\bf k} n}\rangle$ are not gauge invariant under rotation in the composite manifold. Indeed, any  transformation of the kind
\begin{equation}\label{eq:bloch_function_smooth}
    \ket{\widetilde{\psi}_{\mathbf{k}n}} =  \sum_{m=1}^{N_{w}} U_{mn}(\mathbf{k})\ket{\psi_{\mathbf{k}m}} 
\end{equation}
produces equally acceptable Bloch functions, where $U_{mn}(\mathbf{k})$ is a $\mathbf{k}$-periodic unitary matrix rotation in the composite manifold, mixing bands $n$ and $m$ at wave vector $\mathbf{k}$\cite{PhysRevB.56.12847, PhysRevB.65.035109}. 

The $n$-th Wannier function on the ${\bf R}$-th cell is defined as
\begin{equation}
    \ket{\mathbf{R}n} = \frac{1}{\sqrt{N_k^w}} \sum_{\bf k=1}^{N_k^w}\sum_{m=1}^{N_{w}}e^{-i\mathbf{k}\cdot \mathbf{R}}U_{mn}(\mathbf{k})|\psi_{{\bf k}m}\rangle, 
    \label{eq:generalized_wann_func}
\end{equation}
where $N_k^w$ is the number of points in the $\mathbf{k}$-grid used to perform the summation (i.e., the number of electron momentum $\mathbf{k}$-points in the Wannier procedure).  Eq.\eqref{eq:generalized_wann_func} shows that the $N_w$ Wannier functions are not unique as different choices of $U_{mn}(\mathbf{k})$ lead to different Wannier functions with different localization properties. The converse relation of Eq.\eqref{eq:generalized_wann_func} is: 
\begin{equation}
|\psi_{{\bf k}m}\rangle =\frac{1}{\sqrt{N_k^w}}\sum_{\mathbf{R}}\sum_{n=1}^{N_{w}} e^{i\mathbf{k} \cdot \mathbf{R}}U_{nm}^{*}(\mathbf{k})\ket{\mathbf{R}n}.
\label{eq1converse}
\end{equation}

In a topologically trivial insulator, the localization properties
of the Wannier functions are related to the regularity of the periodic part of the Bloch function, $u_{\mathbf{k}n}=e^{-i\mathbf{k} \cdot \mathbf{r}}\psi_{\mathbf{k}n}\sqrt{N_k^w}$
as a function of $\mathbf{k}$ \cite{PhysRev.115.809,PhysRevLett.98.046402}. The more regular the periodic part, the more localized the Wannier functions\cite{PhysRev.115.809,PhysRev.135.A685,PhysRevB.18.4104}.
Exponential localization is obtained if and only if the functions $u_{\mathbf{k}n}$
are analytic in ${\bf k}$ \cite{PhysRev.135.A685,Katznelson}. 

In metals, no such theorem exists and Wannier functions are generally not exponentially localized. However, if a substantial degree of real-space localization in the Wannier functions is achieved, then the associated composite set of bands (obtained after the disentanglement procedure) displays a substantial degree of smoothness in the $\mathbf{k}$-dependence of  $u_{\mathbf{k} n}$. 

A practical way to obtain real-space localized Wannier functions is to use the Maximally Localized Wannier Functions (MLWFs) approach developed in Refs. \cite{PhysRevB.56.12847, PhysRevB.65.035109, RevModPhys.84.1419}. This scheme exploits 
the arbitrariness of the unitary matrix $U_{mn}(\mathbf{k})$ to find the set of Wannier functions for which the spread functional
\begin{equation}\label{eq:SpreadFunctional}
    \Omega = \sum_{n=1}^{N_w} \left[ \langle r^2 \rangle_n - \overline{\mathbf{r}}^2_n \right]
\end{equation}
is minimized. The quantities in Eq.\eqref{eq:SpreadFunctional} are defined as: $\langle r^2 \rangle_n = \bra{\mathbf{0}n}r^2\ket{\mathbf{0}n}$ and $\overline{\mathbf{r}}_n = \bra{\mathbf{0}n}\mathbf{r}\ket{\mathbf{0}n}$. 

Once the spread functional is minimized, a special set of MLWFs and rotation matrix $U_{mn}(\mathbf{k})$ are obtained. The set of Bloch functions obtained from these quantities via the transformation in Eq.\eqref{eq1converse} forms the {\it optimally smooth subspace} (OSS). Many applications of the MLWFs method \cite{RevModPhys.84.1419} have shown that even in the case of metals, localized Wannier functions can be obtained and the OSS transformation leads to smooth periodic parts of the Bloch functions.

\subsection{Operators in the optimally smooth subspace}\label{subsec:IntInOptSmoothSub}

The smoothness of the periodic parts of the Bloch functions guarantees that the matrix elements of  nonsingular operators in the OSS are also smooth. In this work, we focus on operators 
$\mathcal{O}$ that are non-diagonal in the crystal momentum and band indexes, as these are the most time consuming to compute.
The matrix elements of the operator $\mathcal{O}$ over the periodic part of the Bloch functions $u_{\mathbf{k}n}^{PW}$ resulting directly from a density functional theory calculation reads:
\begin{equation}
    \mathcal{O}_{mn}(\mathbf{k}\!+\!\mathbf{q}, \mathbf{k}) = \bra{u_{\mathbf{k}+\mathbf{q}m}^{PW}}\mathcal{O}(\mathbf{q})\ket{u_{\mathbf{k}n}^{PW}}_{\Omega},
    \label{eq:qe_ME}
\end{equation}
where the integration is over the unit cell of volume $\Omega$.
We note that the crystal momenta $\mathbf{k}$ and $\mathbf{k}+\mathbf{q}$ are necessarily points of the $N_{k}^{w}$ grid used in the MLWFs procedure. The matrix element in Eq.\eqref{eq:qe_ME} is typically obtained as an output of a first-principles electronic structure code. The  electron-phonon matrix element  is obtained from Eq. \ref{eq:qe_ME} by choosing ${\cal O}(\mathbf{q})=\frac{\partial v_{KS}}{\partial \mathbf{u}_{\mathbf{q}s}} = e^{-i\mathbf{q}\cdot\mathbf{r}} \frac{\partial V_{KS}}{\partial \mathbf{u}_{\mathbf{q}s}}$ where $V_{KS}$ is the Kohn-Sham potential and $\mathbf{u}_{\mathbf{q}s}=\frac{1}{N_k^w}\sum_{\mathbf{R}_L} e^{-i\mathbf{q}\cdot\mathbf{R}_L} \mathbf{u}_{Ls}$ is the Fourier transform of the phonon displacement $\mathbf{u}_{Ls}$ on the cell $\mathbf{R}_L$, from which it follows that $\frac{\partial}{\partial \mathbf{u}_{\mathbf{q}s}}=\sum_{\mathbf{R}_L} e^{i\mathbf{q}\cdot\mathbf{R}_L} \frac{\partial}{\partial \mathbf{u}_{Ls}}$. The oscillator strength is calculated from Eq.\eqref{eq:qe_ME} in the special case of a
$\mathbf{q}$-independent operator, namely ${\cal O}=e^{i\mathbf{G}\cdot\mathbf{r}}$.

The minimization of the spread functional leads to the unitary transformation $U_{mn}(\mathbf{k})$ identifying the OSS.
Making use of these unitary matrices, the matrix elements of $\mathcal{O}$ in the OSS read
\begin{equation}\label{eq:op_in_optimally_smooth_sub}
    \mathcal{O}_{mn}^{oss}({\mathbf{k}^\prime}, \mathbf{k}) = 
        \sum_{m',n'} U_{mm'}^*(\mathbf{k'})\mathcal{O}_{m'n'}(\mathbf{k}^\prime,\mathbf{k})U_{n'n}(\mathbf{k}),
\end{equation}
where $\mathbf{k}^\prime \equiv \mathbf{k+q}$.

Once the operator in the OSS is known for the crystal momenta $\mathbf{k}$ and $\mathbf{k}^\prime$ belonging to the $N_k^w$ momentum grid, any technique can be used to interpolate it at arbitrary crystal momentum. 
Currently, the standard interpolation method consists of a Fourier interpolation that assumes the periodicity of the operator matrix element in crystal momentum. The goal of this paper is to introduce a different interpolation approach that can also be used for non-periodic matrix elements.

We first review the standard Fourier interpolation approach and then introduce our interpolation scheme.

\subsection{Fourier Wannier interpolation}\label{subsec:IntInOptSmoothSub::FWI}
\begin{figure*}[htp]
    \centering
    \includegraphics[width=\textwidth]{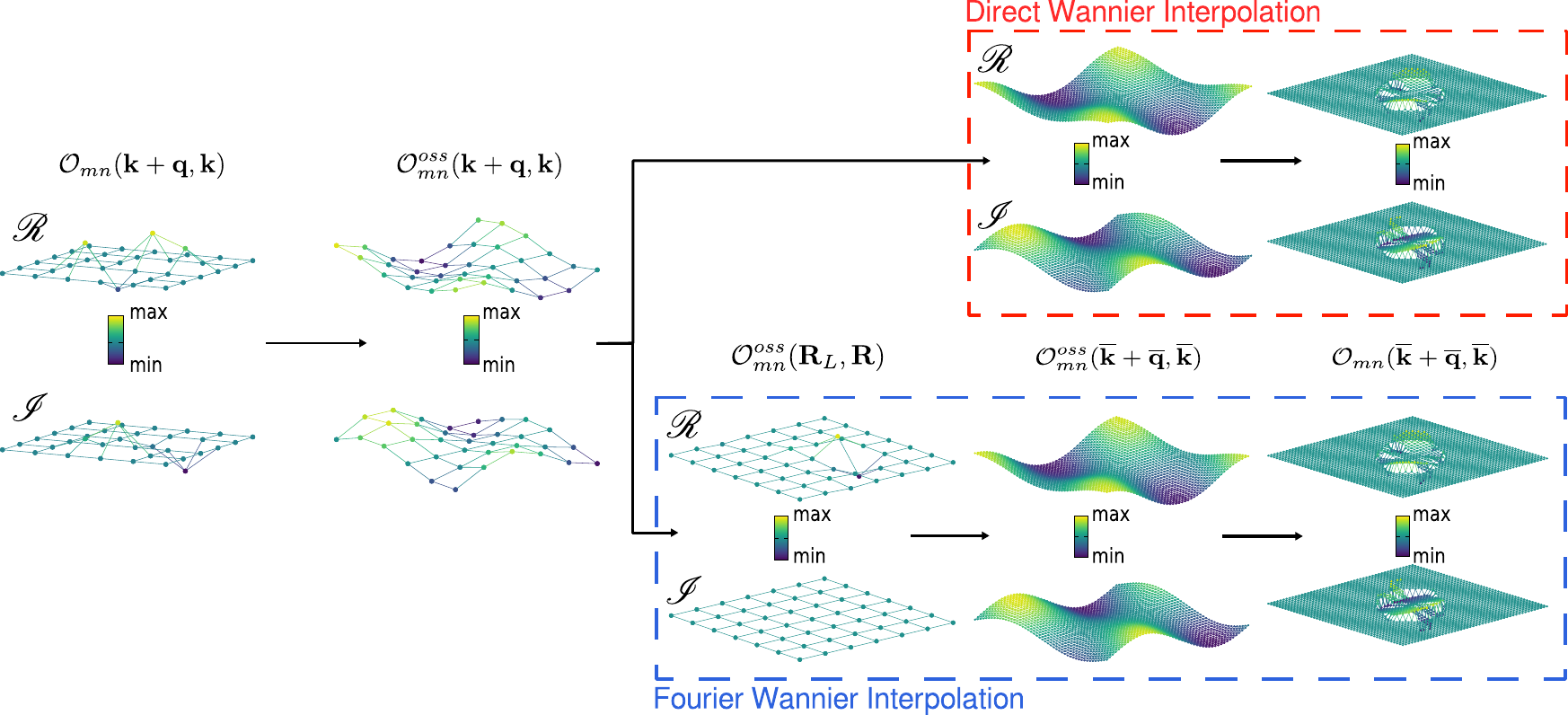}
    \caption{Schematic representation of the steps required in the two interpolation methods. The symbols $\mathscr{R}$ and $\mathscr{I}$ stand for the real and imaginary parts respectively.}
    \label{fig:int-procedure}
\end{figure*}

In the standard Fourier Wannier interpolation (FWI) approach, $\mathcal{O}_{mn}^{oss}({\mathbf{k}^\prime}, \mathbf{k})$ is assumed to be periodic both in $\mathbf{k}$ and
$\mathbf{k}^\prime$. A transformation to Wannier functions basis in the real-space $N_k^w$-~supercell is carried out, namely:
\begin{equation}\label{eq:real_space_Operator_repr}
\begin{split}
    \mathcal{O}_{mn}^{oss}(\mathbf{R_L},\mathbf{R}) &= \bra{\mathbf{0}m}\mathcal{O}(\mathbf{R}_L)\ket{\mathbf{R} n} \\
    &= \frac{1}{(N_k^w)^2}\sum_{\mathbf{k,q}=1}^{N_k^w} e^{-i\mathbf{k}\cdot\mathbf{R}-i\mathbf{q}\cdot\mathbf{R}_L}\mathcal{O}_{mn}^{oss}(\mathbf{k}+\mathbf{q},\mathbf{k}).
\end{split}
\end{equation}
In Eq.\eqref{eq:real_space_Operator_repr}, the vectors $\mathbf{R}$ and $\mathbf{R}_L$ are direct lattice vectors and belong to the real-space $N_k^w$-~supercell. In the case of the deformation potential, ${\cal O}({\bf R}_L)=\frac{\partial V_{KS}}{\partial {\bf u}_{Ls}}$. The matrix elements are then interpolated at any crystal momenta $\overline{\mathbf{k}}$ and $\overline{\mathbf{q}}$ by carrying out the following Fourier interpolation:
\begin{equation}
\label{eq:FWI}
\begin{split}
    \mathcal{O}_{mn}^{oss}(\overline{\mathbf{k}^\prime},\overline{\mathbf{k}}) = 
    \sum_{\mathbf{R},\mathbf{R}_L} e^{i\overline{\mathbf{k}}\cdot\mathbf{R}+i\overline{\mathbf{q}}\cdot\mathbf{R}_L}
    \mathcal{O}_{m'n'}^{oss}(\mathbf{R_L},\mathbf{R}),
\end{split}
\end{equation}
where $\overline{\mathbf{k}^\prime}\equiv\overline{\mathbf{k}}+\overline{\mathbf{q}}$. Finally, the operator in the $u_{\mathbf{k}n}^{PW}$ basis is obtained from the inverse of the transformation in Eq.\eqref{eq:op_in_optimally_smooth_sub}, namely:
\begin{equation}\label{eq:interpolated_op}
    \mathcal{O}_{mn}(\overline{\mathbf{k}^\prime},\overline{\mathbf{k}})=\sum_{m',n'} U_{mm'}(\overline{\mathbf{k}'})\mathcal{O}_{m'n'}^{oss}(\overline{\mathbf{k}^\prime},\overline{\mathbf{k}})U_{n'n}^*(\overline{\mathbf{k}}).
\end{equation}

Fourier interpolation relies on two assumptions: (i) the real-space localization of Wannier functions and (ii) the periodicity of the operator matrix elements.
The first assumption, real space localization of Wannier functions, is related to the smoothness, but in a non-trivial way. Thus, it is not obvious to infer if a  Fourier interpolation over the first Brillouin zone would be more accurate and efficient than a direct multidimensional interpolation of the operator ${\cal O}^{oss}  (\mathbf{k}+\mathbf{q},\mathbf{k})$. 

The second assumption inhibits the application of the Fourier interpolation scheme to non-periodic quantities, such as the oscillator strength. On the contrary, this is possible in the case of multidimensional direct interpolation.

\section{Direct Interpolation in the optimally smooth subspace}\label{sec:IntInOptSmoothSub::DWI}

The Direct Wannier Interpolation (DWI) introduced here is based on the idea that the smooth operator ${\cal O}^{oss}$ can be interpolated via any technique without necessarily using Fourier interpolation. Furthermore, unlike Fourier interpolation, direct interpolation schemes do not require any transformation to real space.

We adopt an algorithm based on successive one-dimensional tridiagonal spline interpolations\cite{NumRecInC}, but other choices are possible.  The interpolation scheme starts again from the matrix elements $\mathcal{O}_{mn}^{oss}(\mathbf{k+q},\mathbf{k})$, defined in Eq.\eqref{eq:op_in_optimally_smooth_sub}, but instead of taking its Fourier transform we perform a direct interpolation in crystal momentum space to obtain $\mathcal{O}_{mn}(\overline{\mathbf{k}}+\overline{\mathbf{q}},\overline{\mathbf{k}})$, i.e. the operator $\mathcal{O}$ at any $\overline{\mathbf{k}}$ and $\overline{\mathbf{q}}$ points of the Brillouin zone.

A visual summary of the steps involved in the two interpolation schemes is displayed in Fig. \ref{fig:int-procedure}.

\subsection{Interpolation of non-periodic matrix elements}

As an example, we consider an operator whose matrix elements $\mathcal{O}_{\mathbf{k}+\mathbf{q},\mathbf{k}}^{nm}(\bf G)$, where $\mathbf{G}$ is a reciprocal lattice vector and both $\mathbf{q}$ and $\mathbf{k}$ are in the first Brillouin zone, are non-periodic in reciprocal space. 
An example of such an operator is the oscillator strength:
\begin{equation}\label{eq:npop_matelem}
    {\cal O}_{\mathbf{k}+\mathbf{q},\mathbf{k}}^{mn}(\mathbf{G})=\rho_{\mathbf{k}+\mathbf{q},\mathbf{k}}^{mn}(\mathbf{G}) = \mel**{u_{\mathbf{k+q}m}^{PW}}{e^{i\mathbf{G}\cdot\mathbf{r}}}{u_{\mathbf{k}n}^{PW}}_{\Omega}.
\end{equation}
 The matrix elements in Eq.\eqref{eq:npop_matelem} are non-periodic in the $\mathbf{q}$ momentum, i.e., ${\cal O}_{\mathbf{k}+\mathbf{q+G'},\mathbf{k}}^{mn}(\mathbf{G}) = {\cal O}_{\mathbf{k}+\mathbf{q},\mathbf{k}}^{mn}(\mathbf{G+G'}) \ne {\cal O}_{\mathbf{k}+\mathbf{q},\mathbf{k}}^{mn}(\mathbf{G})$, limiting the applicability of the FWI.\footnote{{The non-periodicity of the matrix elements is a direct consequence of the \textit{periodic gauge} choice, where $\psi_{\mathbf{k}n} = \psi_{\mathbf{k+G}n}$}} 
 
One possibility could be to extend fictitiously the periodicity of the matrix elements beyond the first BZ, e.g. to a $ 3\times 3\times 3$ reciprocal-space supercell centered in the $\mathbf{G}$-vector of interest. However, this would still result in a systematic error as the matrix elements to be interpolated do not have the periodicity of the chosen supercell, leading to the non suitability of Eqns.\eqref{eq:real_space_Operator_repr} and \eqref{eq:FWI}. Furthermore, the computational cost of this approach is much higher than a FWI performed independently on each unit cell belonging to the supercell. The interpolation on the supercell scales as $(N_k^w)^2\times (N_{uc}^S)^2$, $N_{uc}^S$ being the number of unit cells within the supercell $S$, while the interpolation on the unit cells scales as $(N_k^w)^2\times N_{uc}^S$, a factor $N_{uc}^S$ less. For these reasons, we did not explore this alternative and instead relied on the interpolation on unit cells. 
We show that with the DWI scheme introduced here, we do not introduce any systematic error while maintaining a computational cost comparable to that of the FWI on unit cells.

\begin{figure}[htp]
    \centering
    \includegraphics[width=\linewidth]{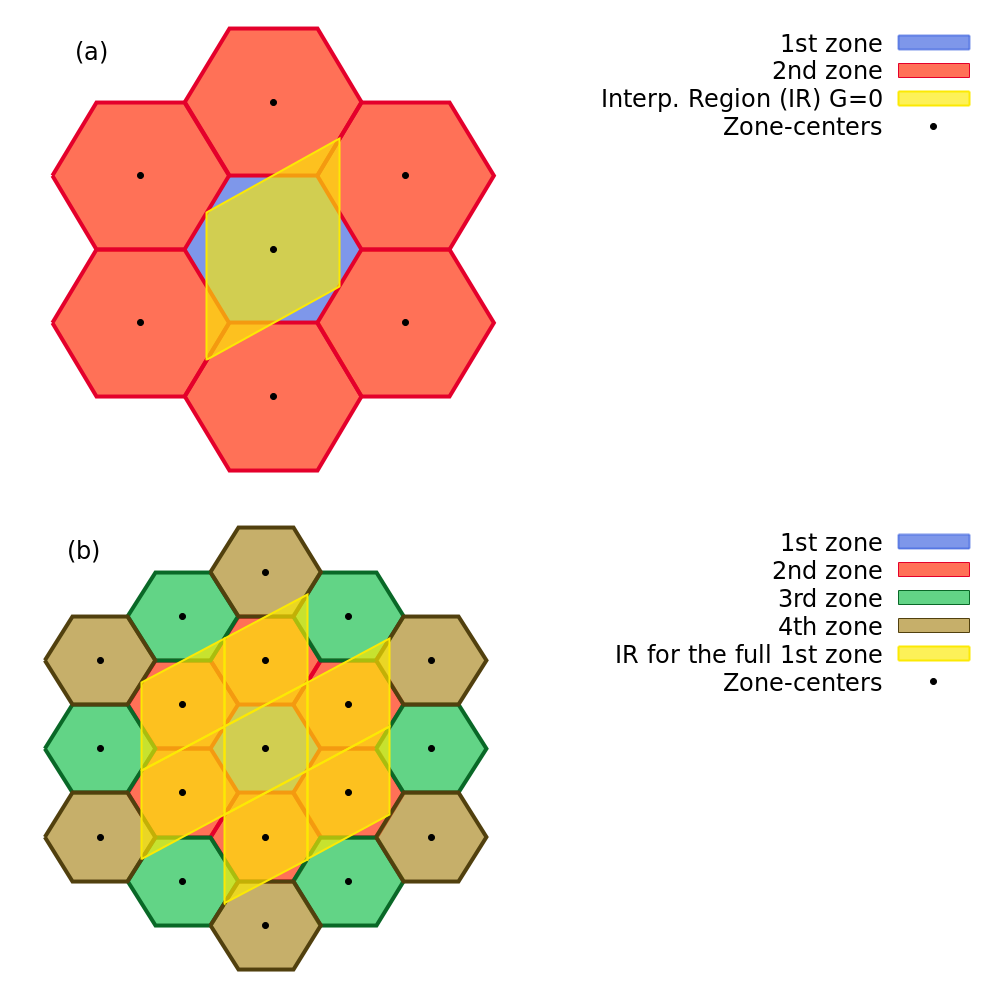}
    \caption{Visual representation of a two-dimensional honeycomb lattice. Panel (a): the blue hexagon is the first zone while the red ones are the second zones i.e. $\mathbf{G}$-shifted BZ. The yellow parallelogram is the region where the interpolation is performed. Panel (b): same as (a) but showing all the parallelograms required to cover the first zone.}
    \label{fig:non-periodic-procedure}
\end{figure}

To illustrate how the interpolation procedure of non-periodic matrix elements works, we consider the hexagonal lattice as an example and, for simplicity, a two-dimensional system. A visual representation of momentum space is reported in Fig. \ref{fig:non-periodic-procedure}. Each hexagon represents one Brillouin zone, translated from the $\mathbf{G}=\mathbf{0}$ one by a $\mathbf{G}$-vector.
For the sake of simplicity, we first focus on the matrix elements interpolation inside the first BZ ($\mathbf{G}=0$). The same procedure can be applied to interpolation on any $\mathbf{G}$-shifted BZ, i.e., any hexagon other than the blue one of Fig. \ref{fig:non-periodic-procedure}. 

To employ the spline-based DWI, as well as for the FWI, we need to know the target operator on a $\Gamma$-centered Monkhorst-Pack grid\cite{Monkhorst_1976}, sampling the yellow parallelogram depicted in panel (a) of Fig. \ref{fig:non-periodic-procedure}. As the figure shows, this requires that the matrix elements outside the first Brillouin zone are known. Once the matrix elements have been computed on the $N^w_k$ grid from first principles on the $\mathbf{G}$-vectors necessary for the construction of the parallelogram, we can perform the interpolation at any point in the yellow parallelogram (interpolation region). Nevertheless, the interpolation region does not cover the full first BZ.

To interpolate the matrix elements throughout the first Brillouin zone, we need to compute them from first principles on a larger region of reciprocal space. That is, we need to cover the first BZ with  $\mathbf{G}$-shifted parallelograms 
(see the interpolation region in Fig. \ref{fig:non-periodic-procedure}, panel (b)). Once we know the matrix elements on all the yellow zones in Fig. \ref{fig:non-periodic-procedure}(b), we can interpolate them on the full first Brillouin zone by using the DWI.

The interpolation of any $\mathbf{G}$-shifted BZ follows the same steps, considering instead the interpolation region centered on the corresponding $\mathbf{G}$-vector.  As it can be understood from Fig. \ref{fig:non-periodic-procedure}(b), a total of seven parallelograms are required to interpolate the full first BZ, two of them being required just for one point at the edges of the first BZ. The corresponding number of $\mathbf{G}$-vectors necessary to fill those parallelograms with the correct points, hence where the first principles calculation must be performed, is nineteen. If one is interested in the whole second zone shell, i.e. the first BZ plus the six surrounding $\mathbf{G}$-shifted BZ, the number of parallelograms is 17 and the number of $\mathbf{G}$-vectors is 35.

In practice, once the operator matrix elements have been calculated from first principles on the chosen supercell of yellow parallelograms, then the interpolation steps are those described in Sec. \ref{subsec:IntInOptSmoothSub}, \ref{subsec:IntInOptSmoothSub::FWI} for the FWI and in Sec. \ref{subsec:IntInOptSmoothSub}, \ref{sec:IntInOptSmoothSub::DWI} for the DWI; schematically displayed in Fig. \ref{fig:int-procedure}. 

To summarize, the interpolation of non-periodic quantities for a generic $\mathbf{G}$-vector is performed as follows:
\begin{itemize}
    \item Construction  of a reciprocal space tessellation formed by a set of suitably chosen reciprocal space unit cells  (yellow parallelograms in panel (b) of Fig. \ref{fig:non-periodic-procedure});
    \item First principles calculation of the target operator on a uniform grid in the part of the reciprocal space identified by the previous tessellation; 
    \item Transformation of the target operator to the optimally smooth subspace via Eq.\eqref{eq:op_in_optimally_smooth_sub},  i.e. ${\cal O}_{\mathbf{k}+\mathbf{q},\mathbf{k}}^{mn}(\mathbf{G}) \rightarrow {\cal O}_{\mathbf{k}+\mathbf{q},\mathbf{k}}^{mn, oss}(\mathbf{G})$;
    \item Interpolation of  ${\cal O}_{\mathbf{k}+\mathbf{q},\mathbf{k}}^{mn, oss}(\mathbf{G})$ using the DWI scheme on an arbitrarily dense grid of points;
    \item Inverse transformation of the interpolated operator via Eq.\eqref{eq:interpolated_op}.
\end{itemize}

As a final note, since the direct interpolation technique is arbitrary in this framework, extrapolation algorithms can also be considered. One could then limit the \textit{ab-initio} calculations to just $\mathbf{G}=0$ and extrapolate the remaining $\mathbf{G}$ of interest. However, this could result in a poor level of accuracy for large $|\mathbf{G}|$.

\section{Applications}\label{sec:Applications}

We now briefly introduce the two operators that will be considered for the comparison of the two interpolation schemes: the electron-phonon coupling and the oscillator strengths. 

\subsection{Electron-phonon matrix elements and related quantities}\label{subsec:Elph}

The electron-phonon coupling can be expressed in a matrix form. Its matrix elements are defined as
\begin{equation}\label{eq:elph_matelem}
    g_{mn}^\nu(\mathbf{k\shortOper{+}q},\mathbf{k}) = \sum_s \frac{\mathbf{e}_{\mathbf{q},\nu}^s}{\sqrt{2M_s\omega_{\mathbf{q},\nu}}}\cdot \mathbf{d}^s_{mn}(\mathbf{k\shortOper{+}q},\mathbf{k}),
\end{equation}
where $\nu$ is the phonon mode index, $m,n$ electronic band indices, $\mathbf{e}_{\mathbf{q},\nu}^s$ are the phonon eigenvectors and $\omega_{\mathbf{q},\nu}$ are the phonon frequencies, both obtained from the diagonalization of the dynamical matrix. The superscript $s$ is a cumulative index labeling the atom in the cell and the cartesian coordinate. $M_s$ is the $s$-th atom mass and $\mathbf{d}^s_{mn}(\mathbf{k\shortOper{+}q},\mathbf{k})$ is the deformation potential. Within density-functional perturbation theory (DFPT)\cite{Baroni_2001}, the deformation potential is obtained as:

\begin{equation}\label{eq:defPot}
    \mathbf{d}^s_{mn}(\mathbf{k\shortOper{+}q},\mathbf{k}) = 
    \mel**{u_{\mathbf{k\shortOper{+}q}m}}
    {\frac{\partial v_{KS}(\mathbf{r})}{\partial \mathbf{u}_{\mathbf{q}s}}}
    {u_{\mathbf{k}n}}_\Omega,
\end{equation}
where $V_{KS}(\mathbf{r})$ is the screened Kohn-Sham potential for which $\frac{\partial V_{KS}}{\partial \mathbf{u}_{\mathbf{q}s}} = e^{i\mathbf{q}\cdot\mathbf{r}} \frac{\partial v_{KS}}{\partial \mathbf{u}_{\mathbf{q}s}}$ and $\mathbf{u}_{\mathbf{q}s}$ is the $s$-th component of the Fourier transform of the atomic displacement $\mathbf{u}_{Ls}$ ($L$ being the index identifying the $L$-th unit cell). The operator $\frac{\partial v_{KS}}{\partial \mathbf{u}_{\mathbf{q}s}}$ is lattice periodic. Hence, the integration can be reduced to one unit cell only.
The interpolation is performed on the deformation potential, Eq.\eqref{eq:defPot}.

Following  Refs. [\onlinecite{PhysRevB.6.2577,PhysRevB.9.4733,PhysRevB.71.064501,Giustino_2017}], the electron-phonon contribution to the phonon linewidth reads: 
\begin{equation}\label{eq:Allen_ph_linewidth}
\begin{split}
    \gamma_{\mathbf{q},\nu} = &\frac{4\pi\omega_{\mathbf{q}\nu}}{N_k} \sum_{\mathbf{k},m,n} \abs{g_{mn}^\nu(\mathbf{k\shortOper{+}q},\mathbf{k})}^2 \delta(\epsilon_{\mathbf{k}n}- \epsilon_F)\\
    &\times \delta(\epsilon_{\mathbf{k\shortOper{+}q}m} - \epsilon_{F}),
\end{split}
\end{equation}
where $\epsilon_{\mathbf{k}n}$ are the Kohn-Sham energies, $\omega_{\mathbf{q}\nu}$ are the phonon frequencies and $N_k$ is the number of electron-momentum points of the interpolated grid. The summation over spin degrees of freedom has already been performed.
It is worthwhile to recall that Eq.\eqref{eq:Allen_ph_linewidth} has a divergent (incorrect) behavior at the zone center for the intraband contribution to the phonon linewidth, as it misses the threshold for Landau damping\cite{PhysRevB.71.064501}.

The phonon linewidth enters in the definition of the mode-resolved electron-phonon coupling constant $\lambda_{\mathbf{q},\nu}$ via the Allen formula \cite{PhysRevB.6.2577,PhysRevB.9.4733} as follows
\begin{equation}\label{eq:mode-res_lambda}
    \lambda_{\mathbf{q},\nu} = \frac{\gamma_{\mathbf{q},\nu}}{2\pi N(\epsilon_F) \omega_{\mathbf{q},\nu}^2},
\end{equation}
where $N(\epsilon_F)$ is the electron density of states per spin at the Fermi level.

From the knowledge of $\lambda_{\mathbf{q},\nu}$ in Eq.\eqref{eq:mode-res_lambda}, we can define the isotropic Eliashberg spectral function $\alpha^2F(\omega)$ Eq.\eqref{eq:EliashbergFunc}, the integrated electron-phonon coupling $\lambda(\omega)$ Eq.\eqref{eq:integ_elph_couplconst} and the averaged electron-phonon coupling, $\lambda$ Eq.\eqref{eq:avg_elph_couplconst}:

\begin{equation}\label{eq:EliashbergFunc}
    \alpha^2F(\omega) = \frac{1}{2N_q}\sum_{\mathbf{q},\nu} \lambda_{\mathbf{q},\nu}\omega_{\mathbf{q},\nu}\delta(\omega-\omega_{\mathbf{q},\nu}),
\end{equation}

\begin{equation}\label{eq:integ_elph_couplconst}
    \lambda(\omega) = 2\int_0^\omega\frac{\alpha^2F(\omega')}{\omega'}d\omega',
\end{equation}

\begin{equation}\label{eq:avg_elph_couplconst}
    \lambda = \frac{1}{N_q} \sum_{\mathbf{q},\nu}\lambda_{\mathbf{q},\nu},
\end{equation}
where $N_q$ is the number of phonon-momentum points of the interpolated grid.

\subsection{Oscillator strength matrix elements}\label{subsec:osc_str}

The oscillator strength matrix elements are defined as 
\begin{equation}\label{eq:osc_str_matelem}
\rho_{\mathbf{k}+\mathbf{q},\mathbf{k}}^{mn}(\mathbf{G}) = \mel**{u_{\mathbf{k+q}m}}{e^{i\mathbf{G}\cdot\mathbf{r}}}{u_{\mathbf{k}n}}_{\Omega}.
\end{equation}

We stress that $\mathbf{q}$ is in the first Brillouin zone and $\mathbf{G}$ is a reciprocal lattice vector. 
The function $\rho_{\mathbf{k}+\mathbf{q},\mathbf{k}}^{mn}(\mathbf{G})$ at any fixed $\mathbf{k}$,$\mathbf{q}$,$m$,$n$ is not periodic in reciprocal space, namely $\rho_{\mathbf{k}+\mathbf{q},\mathbf{k}}^{mn}(\mathbf{G}+\mathbf{G^\prime})\ne \rho_{\mathbf{k}+\mathbf{q},\mathbf{k}}^{mn}(\mathbf{G})$.

We point out that transforming $\rho_{\mathbf{k}+\mathbf{q},\mathbf{k}}^{mn}(\mathbf{G})$ in real space leads to a non localized quantity, meaning that in reciprocal space the oscillator strengths are less smooth than other observables displaying a higher degree of localization in real space. For this reason, we generally expect the interpolation to be less accurate.

Before describing the results of the interpolation scheme for non-periodic matrix elements, we briefly discuss why the interpolation of oscillator strength is relevant. \textit{Ab-initio} calculations of the oscillator strength matrix elements are routinely performed on dense grids in first principles codes. Nevertheless, if the oscillator strength has to be combined in conjunction with other gauge-dependent quantities that require interpolation, such as the electron-phonon coupling in photoluminescence and absorption\cite{doi:10.1021/acs.nanolett.4c00669, PhysRevLett.125.107401, Chan2023}, it is crucial to ensure that both operators' matrix elements are calculated within the same gauge to obtain consistent results. In this context, the strategy proposed here is to Wannier-interpolate both gauge dependent matrix elements starting from the same wavefunction calculation, automatically ensuring that both operators are represented in the same gauge.

\section{Results and discussion}\label{sec:Results}
To benchmark the new interpolation scheme, we apply it to MgB$_2$ and electron-doped single-layer MoS$_2$. 

All calculations are performed within density functional theory. In the case of MoS$_2$, exchange-correlation is treated in the Local Density Approximation (LDA)\cite{LDA_Kohn1965,PhysRevB.23.5048}, using Optimized Norm-Conserving Vanderbilt (ONCV) pseudopotentials\cite{PhysRevB.88.085117} accounting for semi-core electrons. As for MgB$_2$, we approximate the exchange and correlation energy within the generalized gradient approximation (GGA) with the Perdew-Burke-Ernzerhof (PBE)\cite{Perdew_1996} functional, using Martins-Troullier\cite{Troullier-Martins-pseudo1991} Norm-Conserving pseudopotentials.

The computation of the electronic band structure, phonon dispersion and electron-phonon matrix elements on the coarse grid is done by using \textsc{Quantum ESPRESSO}\cite{QE-2009,QE-2017}. The calculation of the Wannier unitary matrices with the MLWFs approach is performed with the \textsc{wannier90} code\cite{Pizzi_2020}. As for the computation of the oscillator strength matrix elements, we used a modified version of the YAMBO code\cite{Sangalli_2019,MARINI20091392}, performing an \textit{ab-initio} calculation of the oscillator strength matrix elements on a regular grid in the reciprocal space region identified by the yellow parallelograms (see Fig.\ref{fig:non-periodic-procedure}).

All the interpolations in the following sections are performed within a modified version of the EPIq code\cite{MARINI2024108950} release.

\subsection{MgB\textsubscript{2}: electron-phonon properties}
\subsubsection{Technical details}
The phonon dispersion and the electron-phonon matrix elements are calculated on a $N^w_k=6\times 6 \times 4$ k-point (and q-point) $\Gamma$-centerd grid. The linear-response\cite{Baroni_2001} calculation is performed using a Methfessel-Paxton smearing\cite{PhysRevB.40.3616} equal to $\SI{0.025}{Ry}$, a kinetic energy cutoff of $\SI{35}{Ry}$ and a $N_k=16\times 16\times 12$ k-point grid in the Brillouin zone for the electronic integration. The lattice parameters are matched with the experimental ones\cite{Shukla_PhysRevLett.90.095506}, namely $a=\SI{5.826}{\bohr}$ ($\SI{3.083}{\angstrom}$) and $c/a=1.142$.

The $N^w_k$ k-points grid is then interpolated to a $N_k^I=80\times 80\times 80$ k-point grid for all panels of Fig. \ref{fig:MgB2_linewidth} to achieve convergence. The dynamical matrices are interpolated to a $N_q^I=20\times 20 \times 20$ in panel (d) of Fig. \ref{fig:MgB2_linewidth} and on 175 points for panels (a)-(c): 50 equally spaced q-point along $\Gamma$-M, K-$\Gamma$ and $\Gamma$-A and 25 along M-K. We used a Gaussian smearing of $\SI{0.01}{Ry}$ to regularize the Dirac delta functions appearing in Eqs.\eqref{eq:Allen_ph_linewidth} and \eqref{eq:EliashbergFunc}. To assess the accuracy of the interpolation schemes,  we also performed a first principles calculation on the same 175 q-points along the high symmetry path and on the $N_q^I$ q-point grid specified above, both using the $N_k^I$ k-point grid and a Gaussian smearing of $\SI{0.01}{Ry}$.

\subsubsection{Results}

\begin{figure}[htp]
    \centering
    \includegraphics[width=\linewidth]{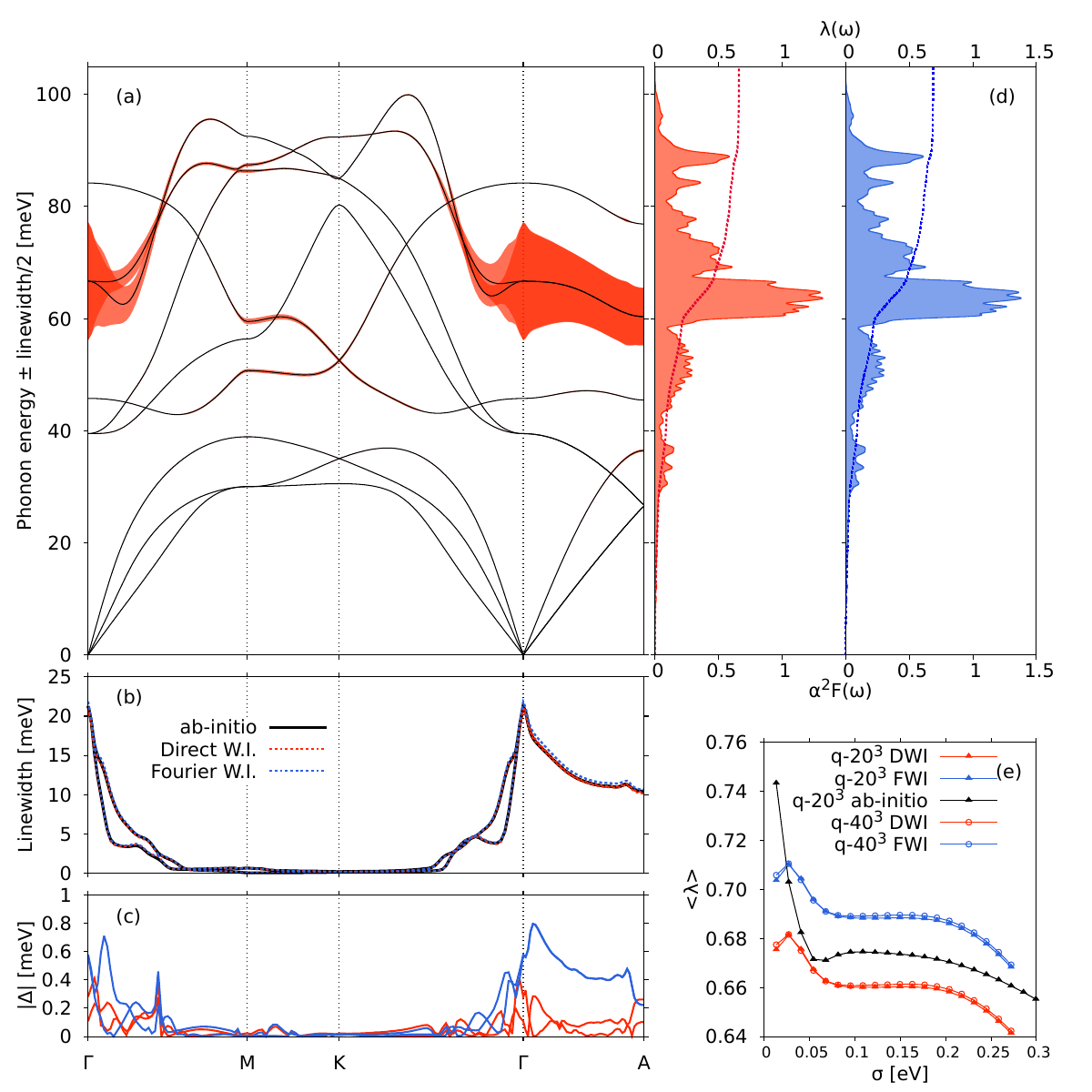}
    \caption{Comparison between the two interpolation schemes for MgB$_2$. Panel (a): black lines represent the phonon dispersion (computed via Fourier interpolation) while the filled red lines are the respective linewidths (Eq.\eqref{eq:Allen_ph_linewidth}) computed using the DWI. Panel (b): linewidth of the E$_{2g}$ modes along the path computed via an \textit{ab-initio} calculation (black), the DWI (red) and the FWI (blue). Panel (c): absolute value of the difference between the \textit{ab-initio} calculation and the interpolation (same color scheme as panel (b)). Panel (d): Eliashberg function $\alpha^2F(\omega)$ (Eq.\eqref{eq:EliashbergFunc}) and the integrated electron-phonon coupling $\lambda(\omega)$ (Eq.\eqref{eq:integ_elph_couplconst}). Panel (e) shows the convergence of $\lambda$ (Eq.\eqref{eq:avg_elph_couplconst}) with the smearing and the q-point grid.}
    \label{fig:MgB2_linewidth}
\end{figure}
In Fig. \ref{fig:MgB2_linewidth}(a), we show the phonon dispersion obtained by diagonalizing the Fourier interpolated dynamical matrices (black lines) as well as the FWHM linewidth associated with each mode (red filled lines). The well-known Kohn anomaly of the E$_{2g}$ modes along the $\Gamma$-$A$ direction and the consequent linewidth increase are well reproduced\cite{Shukla_PhysRevLett.90.095506}. Fig. \ref{fig:MgB2_linewidth}(b) shows a comparison between the linewidth of the E$_{2g}$ modes obtained from 175 independent \textit{ab-initio} calculations (one per q-point, black line) and the two interpolation schemes: in red the direct Wannier interpolation scheme and in blue the Fourier Wannier interpolation one. Both methods are in excellent agreement with the \textit{ab-initio} result. This can be better appreciated in panel (c), where we show the absolute difference between the interpolation and the \textit{ab-initio} calculation. From panel (c), we can see that overall the DWI scheme is, for the present case,  more accurate than the FWI, in particular along $\Gamma A$, where the E$_{2g}$ linewidth is maximal. Fig. \ref{fig:MgB2_linewidth}(d) shows a comparison of the isotropic Eliashberg spectral function $\alpha^2F(\omega)$ (Eq.\eqref{eq:EliashbergFunc}) and of the integrated electron-phonon coupling $\lambda(\omega)$ (Eq.\eqref{eq:integ_elph_couplconst}). Panel (e) shows the convergence with q-points and smearing of the average electron-phonon coupling $\lambda$ (Eq.\eqref{eq:avg_elph_couplconst}). The black line is the result of one independent linear response calculation for each irreducible phonon momentum of the $20^3$ regular grid. Considering the values of average electron-phonon coupling corresponding to a smearing of $\SI{0.010}{Ry}$ $(\simeq \SI{0.136}{eV})$, we get $\lambda_{FWI}=0.688$ for the FWI scheme, $\lambda_{DWI}=0.660$ for the DWI and $\lambda=0.674$ from the \textit{ab-initio} result.
In Tab. \ref{tab:MgB2_avg_ref_comp}, we compare our results with previous calculations in the literature.

\begin{table}[htp]
    \centering
    \begin{tabular*}{\columnwidth}{@{\extracolsep{\fill} } l c c c c c} 
    \toprule
    \\[-0.8em]
    \multicolumn{2}{l}{Reference} & Functional & $\lambda$ & $N_k^I$ & $N_q^I$ \\ [0.5ex]
    \hline
    \\ [-1.5ex]
    \multicolumn{2}{l}{Kong \textit{et al.} (Ref. \onlinecite{PhysRevB.64.020501})} & LDA & 0.87  & $12^2\times 6$ & $6^3$ \\
    
    \multicolumn{2}{l}{Bohnen \textit{et al.} (Ref. \onlinecite{PhysRevLett.86.5771})} & LDA & 0.73  & $36^3$ & $6^3$ \\
    
    \multicolumn{2}{l}{Choi \textit{et al.} (Ref. \onlinecite{Choi2002})} & LDA & 0.73  & $12^2\times 6$ & $18^2\times 12$ \\
    
    \multicolumn{2}{l}{Eiguren \textit{et al.} (Ref. \onlinecite{PhysRevB.78.045124})}& LDA & 0.776  & $40^3$ & $10^3$ \\
    
    \multicolumn{2}{l}{Margine \textit{et al.} (Ref. \onlinecite{Margine_2013})} & LDA & 0.748  & $60^3$ & $30^3$ ($60^3$) \\
    
    \multicolumn{2}{l}{Calandra \textit{et al.} (Ref. \onlinecite{PhysRevB.82.165111})}& PBE & 0.741 & $80^3$ & $20^3$ \\
    
    \multicolumn{2}{l}{Yu \textit{et al.} (Ref. \onlinecite{Yu2024})}& PBE & 0.645 & $60^3$ & $60^3$ \\
    
    This work   & \textit{ab-initio} & PBE & 0.674 & $80^3$ & $20^3$\\
                & FWI       & PBE & 0.688 & $80^3$ & $20^3$\\
                & MSI       & PBE & 0.660 & $80^3$ & $20^3$\\[0.5ex]
    \hline
    \hline
    \end{tabular*}
    \caption{Comparison of the average electron-phonon coupling with values available in the literature.}
    \label{tab:MgB2_avg_ref_comp}
\end{table}

\subsection{MoS\textsubscript{2}: electron-phonon properties}\label{sec:Results:subsec:MoS2:elph}

\subsubsection{Technical Details}

We consider a single-layer electron-doped MoS$_2$ with $n_e=1.15\times 10^{14}$ $e^-$ cm$^{-2}$.
The phonon dispersion and the electron-phonon matrix elements are calculated on the $N^w_k=8\times 8 \times 1$ k-points (and q-points) $\Gamma$-centered grid. The linear-response\cite{Baroni_2001} calculation is performed using a Methfessel-Paxton smearing\cite{PhysRevB.40.3616} of $0.01$ Ry, a kinetic energy cutoff of $80$ Ry and a $N_k=26\times 26\times 1$ k-point grid in the Brillouin zone for the electronic integration. The in-plane lattice parameter is $a=\SI{5.908}{\bohr}$ ($\SI{3.126}{\angstrom}$) and the simulation box,  including vacuum, has $c/a=6.0$.

The $N_k^w$ electron momentum grid is then interpolated to a $N_k^I=192\times 192\times 1$  grid for all panels of Fig. \ref{fig:MoS2_linewidth}. The 
$N_k^w$ phonon momentum grid  is interpolated to a $N_q^I=128\times 128 \times 1$ in panel (d) of Fig. \ref{fig:MoS2_linewidth} and on 125 points for panels (a)-(c): 50 equally spaced q-point along $\Gamma$-M, K-$\Gamma$ and $\Gamma$-A and 25 along M-K. We used a Gaussian smearing of $\SI{0.001}{Ry}$ to regularize the Dirac delta functions in Eqns.\eqref{eq:Allen_ph_linewidth} and \eqref{eq:EliashbergFunc}. To assess the accuracy of the interpolation schemes,  we also performed a first principles calculation on the same 125 q-points along the high symmetry path and on a $N_q^I=96\times 96 \times 1$ q-point grid, both using the $N_k^I$ k-point grid and a Gaussian smearing of $\SI{0.001}{Ry}$.

\subsubsection{Results}

\begin{figure}[htp]
    \centering
    \includegraphics[width=\linewidth]{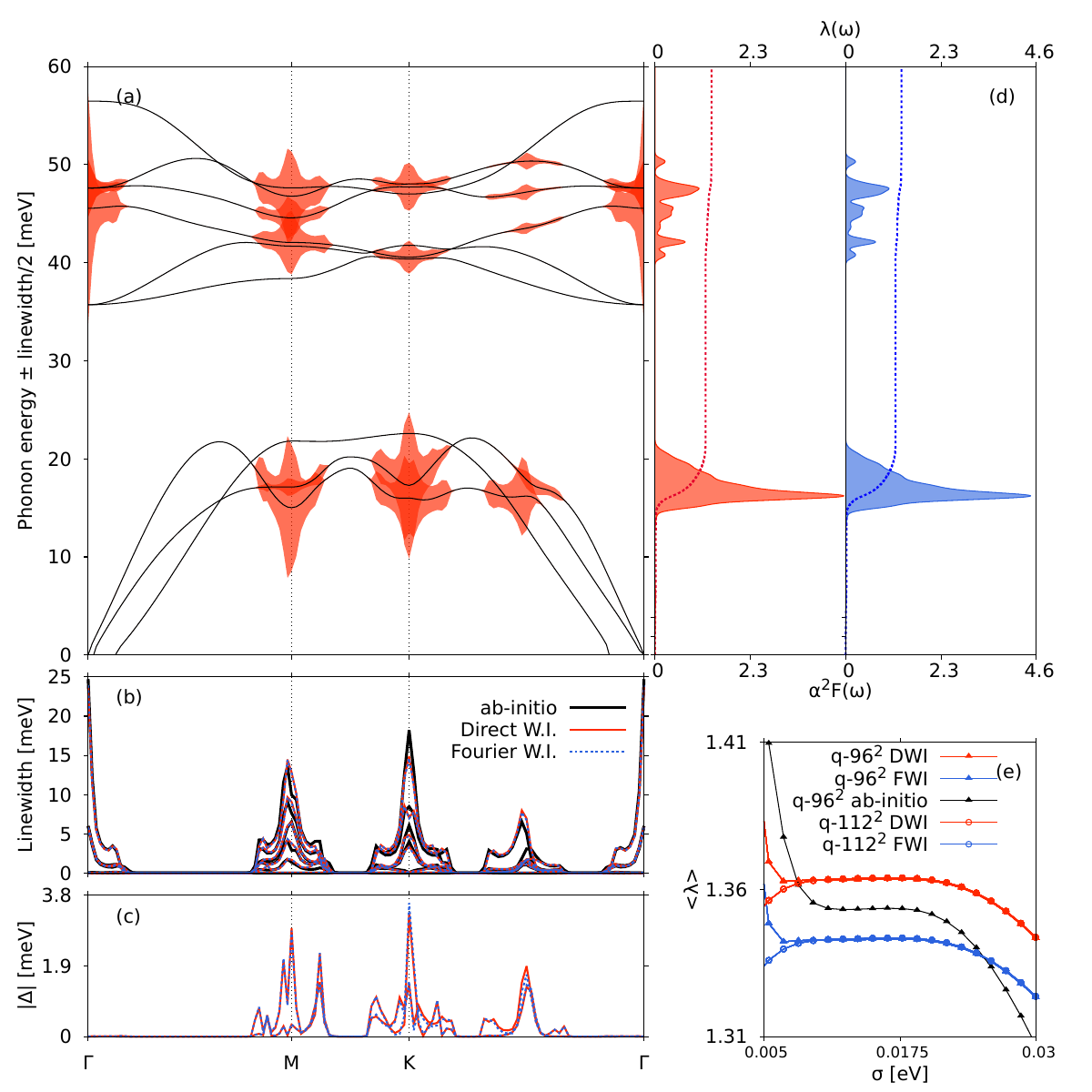}
    \caption{Comparison between the two interpolation schemes for MoS$_2$. Panel (a): black lines represent the phonon dispersion (computed via Fourier interpolation) while the filled red lines are the respective linewidths (Eq.\eqref{eq:Allen_ph_linewidth}) computed using the new interpolation scheme. Panel (b): linewidth of all the modes along the path computed via the new interpolation scheme (red) and the Fourier Wannier interpolation (blue). Panel (c): absolute value of the difference between the first principles calculation and two interpolation schemes. Panel (d): Eliashberg function $\alpha^2F(\omega)$ (Eq.\eqref{eq:EliashbergFunc}) and the integrated electron-phonon coupling $\lambda(\omega)$ (Eq.\eqref{eq:integ_elph_couplconst}). Panel (e) shows the convergence of $\lambda$ (Eq.\eqref{eq:avg_elph_couplconst}) with the smearing and q-point grid.}
    \label{fig:MoS2_linewidth}
\end{figure}

In Fig. \ref{fig:MoS2_linewidth}(a), we show the phonon dispersion obtained by diagonalizing the Fourier interpolated dynamical matrix (black lines) and the FWHM linewidth associated with each mode (red filled lines) calculated with the DWI. Fig. \ref{fig:MoS2_linewidth}(b) shows a comparison between the linewidth of all modes obtained from 125 independent first principles calculations (one per q-point, black line) and the two interpolation schemes; in red the DWI and blue the FWI. Panel (c) shows the absolute difference between the first principles calculation and two interpolation schemes for the two acoustic modes that contribute the most to the Eliashberg function (panel (d)). The two schemes display a similar accuracy in the present case.

Fig. \ref{fig:MoS2_linewidth}(d) shows a comparison of the isotropic Eliashberg spectral function $\alpha^2F(\omega)$ (Eq.\eqref{eq:EliashbergFunc}) and of the integrated electron-phonon coupling $\lambda(\omega)$ (Eq.\eqref{eq:integ_elph_couplconst}). Finally, panel (e) shows the convergence with q-points and smearing of the average electron-phonon coupling $\lambda$ (Eq.\eqref{eq:avg_elph_couplconst}). With a smearing of $\SI{0.001}{Ry}$ $(\simeq \SI{0.0136}{eV})$ the average electron-phonon coupling is $\lambda_{FWI}=1.343$ for the FWI scheme, $\lambda_{DWI}=1.364$ for the DWI and $\lambda=1.353$ from first principles. In this case, the differences between the two interpolation schemes are minimal. A comparison with results available in the literature can be found in Tab. \ref{tab:MoS2_avg_ref_comp}.

\begin{table}[htp]
    \centering
    \begin{tabular*}{\columnwidth}{@{\extracolsep{\fill} } l c c c c c} 
    \toprule
    \\[-0.8em]
    \multicolumn{2}{l}{Reference} & Functional & $\lambda$ & $N_k^I$ & $N_q^I$ \\ [0.5ex]
    \hline
    \\ [-1.5ex]
    \multicolumn{2}{l}{R\"osner \textit{et al.} (Ref. \onlinecite{PhysRevB.90.245105})} & LDA & $\simeq 1.5$  & $32^2$ & $8^2$ \\

    \multicolumn{2}{l}{Ge \textit{et al.} (Ref. \onlinecite{PhysRevB.87.241408})} & LDA & $\simeq 1.2$  & $288^2$ & $4$ \\
    
    This work   & \textit{ab-initio} & LDA & 1.353 & $192^2$ & $96^2$\\
                & FWI       & LDA & 1.343 & $192^2$ & $96^2$\\
                & MSI       & LDA & 1.364 & $192^2$ & $96^2$\\[0.5ex]
    \hline
    \hline
    \end{tabular*}
    \caption{Comparison of the average electron-phonon coupling with values available in the literature.}
    \label{tab:MoS2_avg_ref_comp}
\end{table}

\subsection{MoS$_2$: oscillator strength}\label{sec:Results:subsec:MoS2:oscstr}

We test the interpolation of non-periodic quantities in undoped single-layer MoS$_2$.  We computed the oscillator strength matrix elements on the same $N_k^w=8\times 8\times 1$ grid used in the first principles calculation. For the interpolation of the whole second zone, we calculated the matrix elements for the first zone and the 34 reciprocal lattice $\mathbf{G}$ vectors surrounding it, for a total of 35 $\mathbf{G}$ vectors. We interpolated the matrix elements to a $N_q^I=24\times 24\times 1$ phonon momentum grid and the electron momentum to $\mathbf{k}=\mathbf{K}$, in correspondence to the location of the direct gap. To check the quality of the interpolation, we compared it with the result of a first principles calculation computed on the same $N_q^I$ grid.

\begin{figure}[htp]
    \centering
    \includegraphics[width=\linewidth]{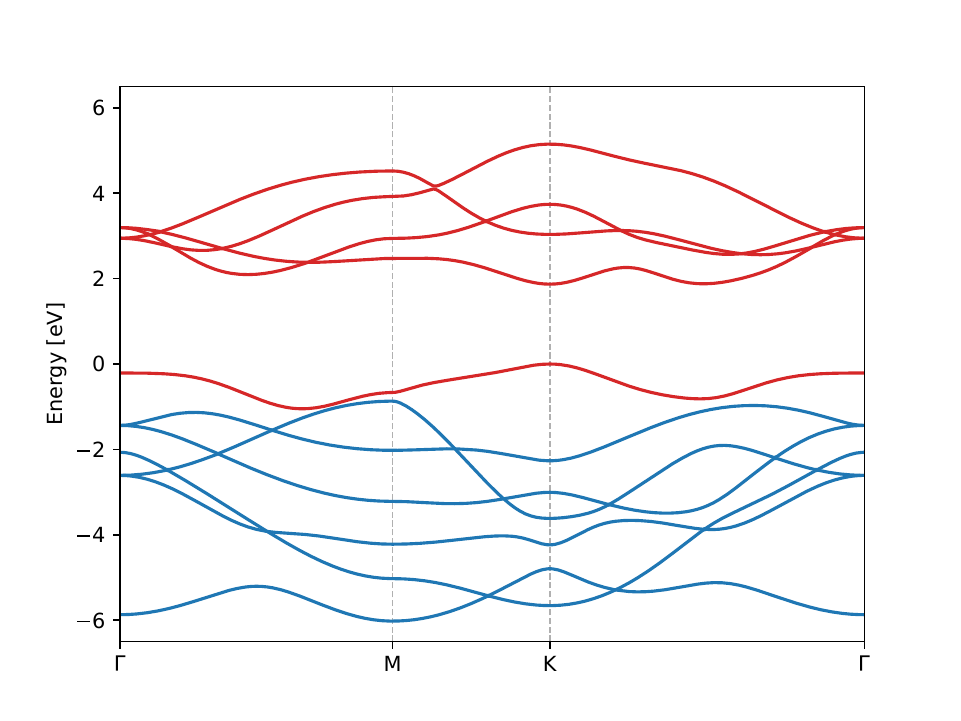}
    \caption{Band structure of MoS$_2$ along the high-symmetry path of the two-dimensional hexagonal Brillouin zone. The red bands are those used in the Wannierization. The maximum energy of the valence manifold is set to zero.}
\label{fig:MoS2_bands}
\end{figure}

\begin{figure}[htp]
    \centering
    \includegraphics[width=\linewidth]{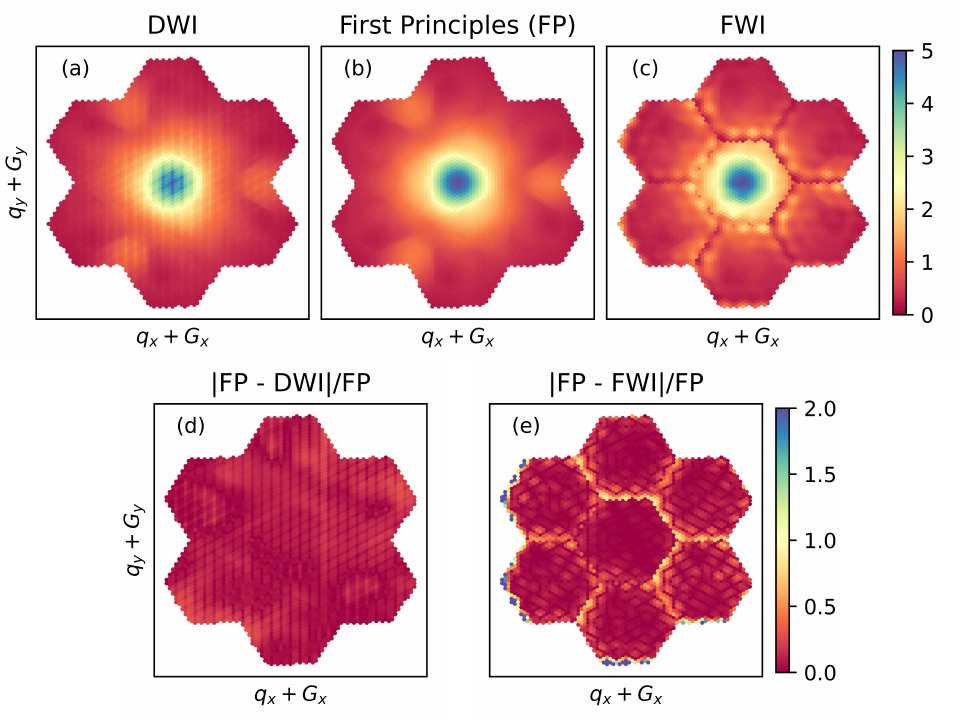}
    \caption{Comparison between the interpolation methods and first principles calculation for Eq.\eqref{eq:osc_str_matelem-tvac}. In the bottom panels, the relative error is plotted. A relative error of $1$ means $100\%$ error with respect to the first principles result.}
    \label{fig:MoS2_osc_strength_tvac}
\end{figure}

In principle, the \textit{ab-initio} calculation on the dense grid is in a different gauge from that of the coarse grid used for the interpolation. We must consider a gauge independent quantity to ensure the comparison is meaningful. This is done by considering the square modulus of the matrix elements relative to an isolated set of bands. By looking at the band structure of single-layer doped MoS$_2$ (Fig. \ref{fig:MoS2_bands}), the top valence band is isolated from all other bands while the bottom conduction band is not. For this reason, in Eq.\eqref{eq:osc_str_matelem-tvac}, we considered the matrix elements connecting the top valence band with all the available conduction bands within the energy window chosen in the disentanglement that cross with the bottom conduction band: 
\begin{equation}\label{eq:osc_str_matelem-tvac}
    |\rho_{\mathbf{k=K},\mathbf{q}}^{tv,cc}(\mathbf{G})|^2 = \sum_{m,n=tv,cc} |\rho_{\mathbf{k=K},\mathbf{q}}^{mn}(\mathbf{G})|^2,
\end{equation}
where $tv$ and $cc$ stand for the top valence and the conduction bands in the sense explained above. The electronic band structure and the selected bands considered in Eq.\eqref{eq:osc_str_matelem-tvac} are displayed in Fig. \ref{fig:MoS2_bands}. In Fig. \ref{fig:MoS2_osc_strength_tvac}, we show the result of the interpolation obtained with both methods and a comparison with the first principles result. Panels (d) and (e) show, respectively, the relative difference (normalized to the first principles result) between the interpolation results and the first principles calculations. To avoid spurious contributions to the relative error, we set it to zero on the points where the first principles value is below a cutoff of $10^{-4}$. From the figure, it is clear that our method can better capture the shape of the oscillator strength, while FWI struggles, especially at zone borders where the relative difference with the first principles result is well above 1.0 (i.e. well above $100\%$ error).

\section{Conclusions}\label{sec:Conclusions}

In this paper, we explored a new strategy for the interpolation of operator matrix elements in the optimally smooth subspace within the Maximally Localized Wannier Functions framework. We demonstrated that the direct interpolation scheme based on successive one-dimensional tridiagonal splines constitutes a valid alternative to the standard Fourier interpolation for Brillouin zone periodic quantities both in terms of quality and computational scalability. Furthermore, by exploiting direct interpolation, we devised an algorithm to tackle the interpolation of non-periodic matrix elements in reciprocal space, overcoming a fundamental limitation of Fourier interpolation. Our approach enables treating periodic and non-periodic quantities in a unified framework. Notable applications include a gauge consistent treatment of electron-phonon and electron-electron interactions, crucial for real-time dynamics and the correct description of equilibrium and non-equilibrium spectroscopic quantities. 

Direct interpolation in the optimally smooth subspace has two additional advantages compared with Fourier interpolation. The first one is that it can be easily employed for periodic and non-periodic tensors of any order without the need for any information on the position of the Wannier centers in real space. 

Secondly, the direct interpolation method is not bounded to the restriction of having equal electron momentum $\mathbf{k}$ and phonon momentum $\mathbf{q}$ coarse grids on which the first-principles calculation of the operator matrix elements is carried out, as it happens in the Fourier case (see the discussion in Fig. 1 in Ref. [\onlinecite{PhysRevB.82.165111}]). Avoiding the real-space step in the direct interpolation allows us to relax the condition to commensurate $\mathbf{k}$ and $\mathbf{q}$ grids. In this way, we can perform the Wannierization on much finer grids, hence increasing the precision of the electronic structure while still keeping the phonon momentum grid at reasonable scales.
This remains true even in the case of higher order tensors.

Finally, a remarkable result of the direct interpolation scheme is that any interpolation method, not limited to any class, can be employed. Hence, the choice can be made depending on the desired level of accuracy and the computational cost.

\section{Acknowledgements}
Funded by the European Union (ERC, DELIGHT, 101052708). Views and opinions expressed are however those of the author(s) only and do not necessarily reflect those of the European Union or the European Research Council. Neither the European Union nor the granting authority can be held responsible for them. We acknowledge the CINECA award under the ISCRA initiative. We acknowledge EUROHPC (EHPC-REG-2024R01-061), for the availability of high performance computing resources and support. M.C. acknowledges fruitful discussions with Francesco Mauri.

\bibliography{bibliography}
\end{document}